# VIDEO DATA VISUALIZATION SYSTEM: SEMANTIC CLASSIFICATION AND PERSONALIZATION

Jamel SLIMI[1] and Anis Ben AMMAR[1] and Adel M. ALIMI[2]
[1]ReGIM (Research Group on Intelligent Machine)
[2]National Engineering School of Sfax, Tunisia
National Engineering School of Sfax University of sfax, Tunisia

{jamel.slimi, anis.benammar, adel.alimi}@ieee.org

## ABSTRACT

*We present in this paper an intelligent video data visualization tool, based on semantic classification, for retrieving and exploring a large scale corpus of videos. Our work is based on semantic classification resulting from semantic analysis of video. The obtained classes will be projected in the visualization space. The graph is represented by nodes and edges, the nodes are the keyframes of video documents and the edges are the relation between documents and the classes of documents. Finally, we construct the user's profile, based on the interaction with the system, to render the system more adequate to its preferences.*

## KEYWORDS

*Video, Semantic Concept Extraction, Semantic Classification, data Visualization, Personalization.*

## 1. INTRODUCTION

Through the technological progress in video production and the availability of cheap digital cameras, the size of video collections is growing tremendously in the personnel database or in the database of dedicated Web sites such as "YouTube", "FaceBook", "DailyMotion", etc [1]. Hence, effective exploration and simple navigation in these collections become an urgent need.

Data visualization is still a relatively active field of research. Considering that vision is the most developed sense in humans. It represents together with the brain psycho-visual vision and cognitive capacities, a privileged tool to analyze one's environment. Visualization research can be defined by the techniques that allow transcribing or modelling complex phenomena or structures, in order to manipulate and analyze them in visual forms.

Video data are very important sources of information. Owing to the fact that they cover most of the daily events, they satisfy the extremely various requests of the public (political, cultural, sport...) [2]. For these reasons, the video data are observed and analyzed by researchers. Therefore, research fields focus more and more on video data processing such as video indexing, concepts extraction, classification....

In recent years, there has been an interest in another type of challenge which is the exploration and navigation in the video documents corpora. Despite the progress in video basic treatments, automatic video data visualization is still suffering from the following problems.

The first problem resides on the semantic analysis of video data. For a good exploration and navigation, visualization system must understand the basic semantics of the documents. Existing video exploration systems can only support the services based on low-level visual features [3]. However using only low-level descriptors (form, texture, and colour) limits the ability of the system. For this, we based our work on high level concepts to improve the quality of the video data visualization system.

The second problem is how to represent the whole of the documents in the visualization space? The visualization system is limited by the visualization surface. So, video collection (e.g. thousands of hours of video) causes an organization problem of the keyframes in the visualization space.

The third problem is how to personalize the visualization system behind receiving the user's input? Before the user's interaction with the system to express his/her demand, the system can only present a general overview of the corpus. This general overview can neither present the details of the corpus nor response to the user preferences efficiently. For that, after the user input, the system must reorganize the documents in an allowing way to disclose the details related to the user's input.

Based on the above elucidation, we propose a tool to help the user to explore a large corpus of video document. The idea is to create a generic framework to fulfil various tasks in video data visualization. Various tasks consist to the video concepts extraction, video data semantic classification, video data organization and video data visualization system personalization.
The first step is devoted to the construction of the video data concepts vectors. The concepts vector is composed of two objects which are: the keyframe that represents the document and the whole concepts that describes it.

The second step is the semantic classification of the video data. The classification is resulting from semantic similarity distance between concepts vectors describing video data.
The third step is the organization of the corpus through the projection of the classes in the visualization space.

The last step is the creation of the user's profiles via his/her interaction with the system to render the video data organization more adequate to his/her preferences.

Compared to existing work, our contributions incorporate the following parts. First of all we treat the various types of video documents. Unfortunately, the existing systems treats just one type of video such as news, sports, etc. Secondly, we use high level concepts for the description of the video sequences which has an important role in improving the semantic classification. Thirdly, the visualization graph is represented in the form of a network based on biological neuronal metaphor. This representation facilitates much navigation in the corpus of documents. Finally and essentially, we integrate a personalization module in order to accelerate the access process to the video documents. Personalization aims at representing the documents according to the preferences and behaviour of the user.

This paper is organized as follows. Section 2 introduces a short state of the art on video visualization system. Then, in section 3, we describe our framework. From sections 4 to 7 we present the different tasks of the visualization processes. We, then, provide conclusions in Section 8 in which perspectives are given for future improvements.

## 2. PROBLEM POSITION AND RELATED WORKS

Due to the large amounts of multimedia data generates every day, an exploration and navigation interface is becoming an increasing need. So data visualization represents a very active field of research in the last years. Data visualization system attempts to exploit human visual processing system innate ability to allow a user to perceive patterns, such as groups and outliers, within large data sets in a more efficient manner than examination of the raw data.

The data visualization system interfaces permit the users to communicate with information in the objective to finalize well defined tasks. The research tasks in interfaces field aim at defining software methods, models and components for the creation of effective interfaces. Effectiveness means the facility of training and use. Indeed, system interface plays a crucial role in the development of interactive applications. Its ease of use can sometimes be a significant criterion in the evaluation of the application. In addition, a good application which is represented by no adequate interface can be judged as an unsuccessful application. In fact, the judgment of the success of the data visualization interface resorts to the user expectations. So, interface must satisfy the needs and the preferences of the user. However users do not have common perceptions and choices. For that, the adaptability of the data visualization system interface becomes an urgent need.

Various approaches exist for data visualization to help users explore in database and find interesting documents intuitively. Proposed approaches are coupled with a variety of systems such as multimedia retrieval engines [4] or annotation tools [5, 6].

Historically, visualization systems are interested in rearrangement of the research results in textual database to facilitate navigation in the retrieved documents. Representations are based on the similarity between the key words describing the documents [7]. Recently, there has been several display systems treating the multi-media documents [8]. What follows is a non exhaustive study of some systems followed by a comparative study. A comprehensive review on video data visualization system can be found in [9].

Visualization systems based on timeline slider [10, 11, and 12] are the most common and easiest way to get a quick overview of video content. The major disadvantages of these tools are the lack of scalability for long documents and the unpleasant aspect due to the poor visual feedback.
Campanella et al. [5, 6, and 13] propose a data visualization system, FutureViewer, to explore and annotate video sequences. The user can interactively visualize classes and annotate them and can also graphically explore how the basic segments are distributed in the feature space. Each video document is considered as a sequence of shots. The feature space is displayed in a 2D Cartesian plane, where each axis corresponds to one feature type selected by the user and each shot is represented by a little square filled by the dominant colour of the shot.

In the Intelligent Sensory Information Systems team of Amsterdam's University, the MediaMill video search engine [14, 15 and 16] proposes four browsing tools for video data. First, Galaxy Browser uses 2D similarity-based visualization of keyframes where the user can annotate a large collection of images. This system can improve the efficiency of the relevance feedback process by mixing relevance feedback and 2D visualization [17, 18]. Second, Sphere Browser, represents a novel interface for searching through semantic space using conceptual similarity. This is due the classification shots with a similar conceptual index together into threads. The Sphere Browser shows the time-line of the current video on the horizontal axis, and for each shot from the video it displays the relevant threads on the vertical axis. Third, Cross Browser, uses a linear ordering to ranking video data. The vertical axis is related to a selected concept. The horizontal one is used to visualize video program in a time-line from which a keyframes is selected. The last, Rotor browser, is an extension of the Cross Browser using more than two axes.

In [19], the proposed System, Table of Video Contents (TOVC), presents a framework for multimedia content visualization and navigation. This tool is adapted for structured programs as news, magazines or sport events. Connected with a player, users can have practical exploration through a global view. The visualization shows the video structure and gives almost instantaneously relevant information such as sequence number, relative durations, dominant colours. . . .

In [20], the authors propose an "instant video browsing" which divides every video into as many parts of equal length, as there are video windows opened on the screen. This tool offers two different views for browsing the video data: the parallel view and the tree view. The limit of this tool resides in the segmentation and the indexing of video data.

In [21, 22, and 23], the authors present initially a simple tool for storing images based on their colour similarities. Just using the colour descriptor decrease the system performance. The advantage of this tool is the organisation of the visualization space based on 3D cylinder form. Secondly, they present a video browsing tool that supports users at the task of interactive search and navigation. This tool is based in 3D user interface organised in the form of Thumbnail Ring arranged by colour similarity.

Data clustering and intuitive representation are the two fundamental ideas developed over all the proposed system. In fact, classification, whatever the inducted method or process, provides an overview of all the data to be acceded. Visualization based on common and intuitive idea helps the user to target the needed information. The major disadvantages of those tools are the treatment of one type of video data (news or sport sequences) and essentially not having explored the user's attention models to enable more effective data visualization.

Our approach covers all the tasks of the visualization process: from indexing to the organization of the video collection in the visualization space. The advantage of the proposed approach consists to (1) the use of high-level features extraction (Visual features, Text features, Audio features) for the description of the video data; (2) The computation of the semantic similarity between the video data based on the correlation between the concepts vectors describing the video collection; (3) the use of metaphor inspired from the biological neuron function that allows a network representation in the visualization space composed by nodes and edges; (4) the integration of a user model in the visualization system allowing a representation adequate to the user preferences. In result, our tool allows a simple, intuitive and efficient navigation in large-scale video corpus.

In the following section we will detail our framework to represent and visualize large scale video corpora.

## 3. FRAMEWORK DESCRIPTION

In this section, we present an overview of our plate form architecture. Our framework is composed of four parts. The starting point of the data visualization system is a collection of video data. This collection contains various type of video (film, sport, news…).

The first step consists of high-level concepts extraction which is based on video text extraction, visual feature extraction and audio feature extraction. These concepts allow constructing the concept vectors describing the video data.

The second step is dedicated to the construction of the similarity matrices. We calculate the semantic similarity distance between all the concept vectors of video collection. The similarity matrices are used to group elements and to explicit the relation between elements.

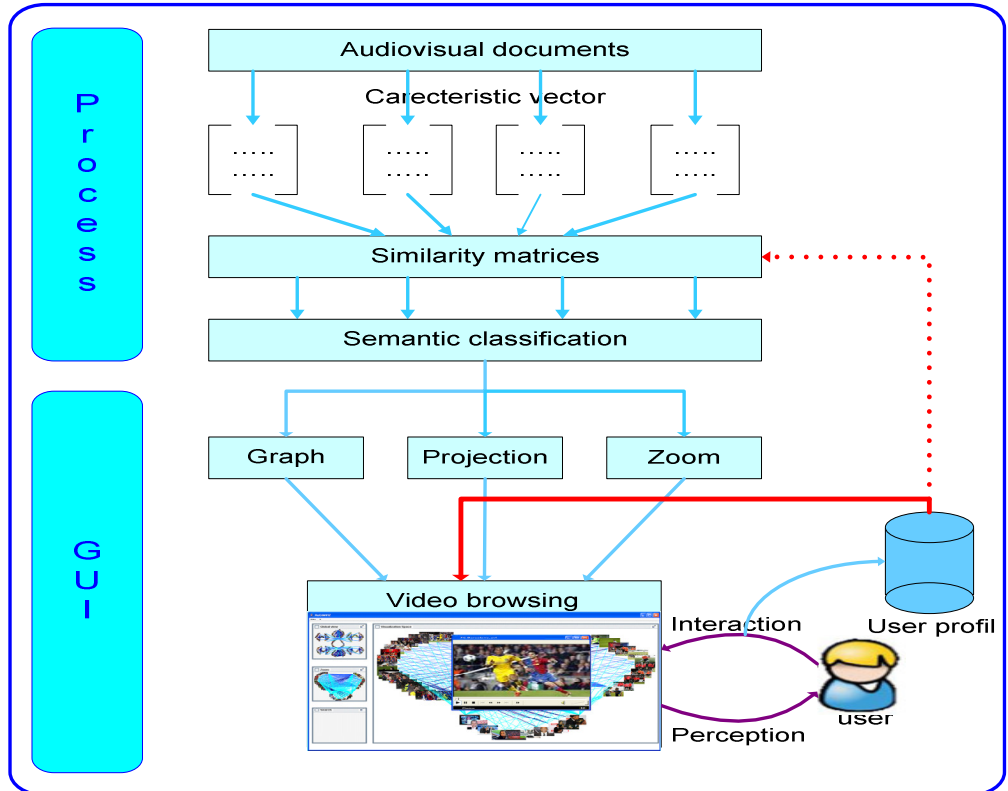

Figure1. Framework Description

The third step is devoted to the projection of the obtained classes (groups) on the visualization space. The organisation of video collection is based on network representation. Key frames represent the nodes of the representation and the edges represent the relations between elements.
The last step is dedicated to the personalization of the visualization system. The visualization space does not make it possible to post the totality of the documents composing the corpus. Personalization is expressed by the creation of the users' profiles via their interaction with the system. These users' profiles will influence directly the visualization space and/or the semantic distances calculation to post only the significant documents for the user.

The approach adopted for the realization of the visualization system has several advantages. Firstly, we treat various types of video data. Also, the use of the high level concepts improves semantic classification. Visualization in the form of a network explains the relations inter-documents and facilitates the task of navigation in the corpus. The recourse to the user preferences modelling allow a posting according to his/her needs that gives a fast and effective access to the desired document.

## 4. INDEXING PROCESS

The video data collection used in our work, in the tests phase or for the development phase, is extracted from TREC Video Retrieval Company (data base 2010). Several treatments are applied to the data collection before being used. We based the multimodal indexing process on text concepts extraction, visual feature extraction and audio feature extraction. The whole of the extracted concepts are used to construct the concept vectors of video data. To start the concepts extraction process video sequence must be divided on subsequence (shot). The shot consist of one or more related frames that represent a continuous action in time and space [24]. For ease of use, a shot is represented by a key frame. Generally, the key frame is the central frame of a shot [25]. Once, the video is segmented, we start the concept extraction process. When considering video, the common approach is to take a camera shot as the pattern of interest. Such a pattern can then be described using text features, audio features, visual features, and their combinations [26] (Figure 2).

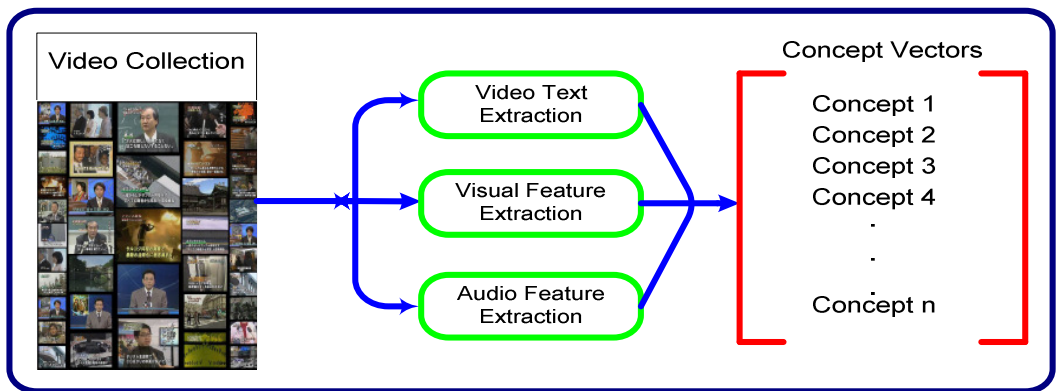

Figure2. Concepts Extraction

### 4.1. Video Text Extraction

A text embedded in video data contains large qualities of useful information. The first step of text concept extraction consists to splitting the video data on frames. The second step consists of the recognition of the embedded text in the frame. Video text extraction involves detection, localization, enhancement and recognition of the textual content in the video frame [27]. To segment the video data on frame, we have used the Java Media Framework package (JMF). The input of this step is the shot containing text. At the rate of one frame per second, the shot is framed into images. The obtained images are sent to an OCR for the text recognition.

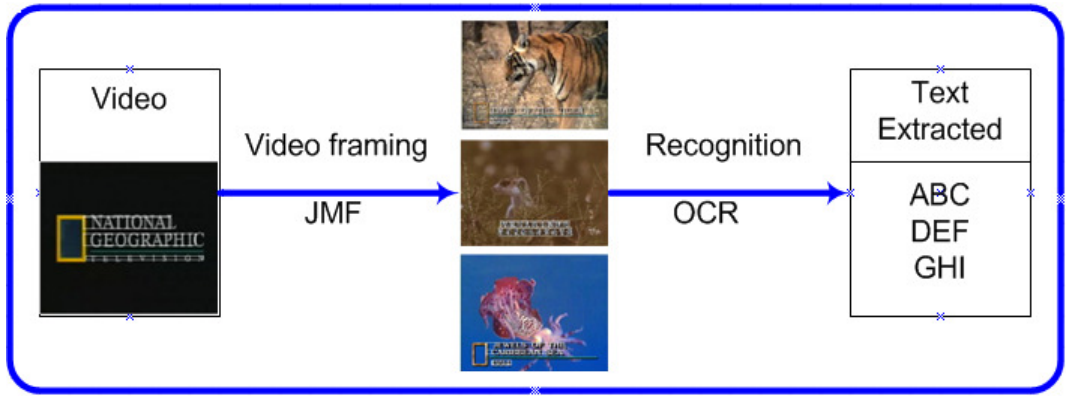

Figure3. Video Text Extraction

## 4.2. Visual Feature Extraction

The aim of the visual feature extraction is to improve the video analysis. There are many visual features [28, 29, 30, and 31]. We can divide visual descriptors into two main groups: General information descriptors and Specific domain information descriptors. The general information descriptor contains low level descriptors which give a description about colour, shape, regions, textures and motion. The specific domain information descriptor gives information about objects and events in the shot.

*Colour*: the key frame is an array of pixels (picture elements) where each pixel has a colour. Colour histogram is the representation of the distribution of colours in the key frame [32]. HSV space colour is defined as follow: Hue is the dominant wavelength in the color spectrum, Saturation is a measure for the amount of white in the spectrum and Volume is a measure for the brightness or intensity of the colour. Colour describing is based on the following tools: Dominant Colour Descriptor (DCD), Scalable Colour Descriptor (SCD), Colour Structure Descriptor (CSD) and Colour Layout Descriptor (CLD).

*Texture*: is an important point for keyframe describing. The texture descriptor observes the region homogeneity and the histograms of these region borders. Descriptors are represented by: Homogeneous Texture Descriptor (HTD), Texture Browsing Descriptor (TBD) and Edge Histogram Descriptor (EHD).

*Shape*: Alternatively, a keyframe can be segmented by grouping pixels in the image based on a homogeneity criterion on color, texture, or both [33, 34, and 35], or by connecting edge lines [36]. Shape contains important semantic information due to human's ability to recognize objects through their shape. These descriptors describe regions, contours and shapes for 2D images. The shape descriptors are the following ones: Region-based Shape Descriptor (RSD), Contour-based Shape Descriptor (CSD) and 3-D Shape Descriptor (3-D SD).

*Region*: the Scale-Invariant Feature Transform (SIFT) descriptor measures the region around a key point and describes each region using an edge orientation histogram. SIFT descriptor is the best performing interest region descriptor. SIFT describe the local shape of the interest region using edge histograms [37]. The interest region is divided into a 4x4 grid and every sector has its own edge direction histogram. The grid is aligned with the dominant direction of the edges in the interest region to make the descriptor rotation invariant.

### 4.3. Audio Feature Extraction

The text data derived from audio analysis is very important to describe video data. For the audio feature extraction we use the Automatic Speech Recognition (ASR). The input of the system is the video shot and the output is a text describing the speech included in the video data. The use of the audio feature improves the video data description.

### 4.4. Supervised Learners

The Concept detection in video data is considered as a pattern recognition problem. For a pattern x and a shot i, the aim is to obtain a probability measure indicating if the semantic concept $C_x$ exists in i. for this, we use the support vector machines (SVM) framework [38]. Also, we use the LIBSVM implementation (for more detail sees http://www.csie.ntu.edu.tw) with radial function and probabilistic output. The SVM yield a probability measure $p(C_j|x_i)$, which we use to rank and to combine concept detection result.

## 5. SEMANTIC CLASSIFICATION

Semantic classification is the most important step in data visualization process. The aim of the classification is to group video data according to their semantic concepts. These groups facilitate the knowledge extraction task from the video data collection. Semantic classification is based on video semantic Extraction. In this section, we present, firstly, the video data representation model, after that we present our experiments about video data concept extraction and the concept correlation, finally we present the similarity distance computation between the concept vectors describing the set of video data composing the collection.

### 5.1. Video Data Representation Model

A video data is represented by two objects. The first is the keyframe and the second is the concept vector which is composed by the high-level concepts extracted from video data. The keyframe is the window representing the document in the visualization space. All the modifications such as insertion, suppression, increase or decrease in size or the change of place will be applied to this object. The keyframe is obtained by the segmentation of the video data. For the document description, we use high level concepts. These concepts are extracted from text, audio and visual constituting the video data. This is done a semantic description of the video. Based on this semantic description we establish the phase of the similarity distances computation.

### 5.2. Concepts Extraction

The number of high-level concepts that was used in our work is 130 concepts (Figure 4). Each document is characterized by a set of semantic concepts. These concepts will constitute the video data concept vectors. The objective of the semantic concepts is to improve semantic classification. Thus, we can facilitate the task of browsing the large scale corpora and accelerate the access information task. The main objective of our system is to provide automatic content analysis and an intelligent tool for retrieval and browsing a video data collection. Firstly, we extract features from the keyframe and then label each keyframe based on corresponding features. For example, if three features are used (colour, motion and audio), each keyframe has at least three labels.

| TV10_ID | LSCOM_Name |
|---|---|
| 001 | Actor |
| 002 | Adult |
| 003 | Airplane |
| 004 | Airplane_Flying |
| 005 | Anchorperson |
| 006 | Animal |
| 007 | Asian_People |
| 008 | Athlete |
| 009 | Basketball |
| 010 | Beach |
| 011 | Beards |
| 012 | Bicycles |
| 013 | Bicycling |
| 014 | Birds |
| 015 | Boat_Ship |
| 016 | Boy |

```
</videoFeatureExtractionFeatureResult>
<videoFeatureExtractionFeatureResult fNum="130">
    <item seqNum="1" shotId="shot10028_1"/>
    <item seqNum="2" shotId="shot10028_2"/>
    <item seqNum="3" shotId="shot10028_3"/>
    <item seqNum="4" shotId="shot10028_4"/>
    <item seqNum="5" shotId="shot10028_5"/>
    <item seqNum="6" shotId="shot10028_6"/>
    <item seqNum="7" shotId="shot10028_7"/>
    <item seqNum="8" shotId="shot10028_8"/>
    <item seqNum="9" shotId="shot4781_32"/>
    <item seqNum="10" shotId="shot4781_33"/>
    <item seqNum="11" shotId="shot4781_34"/>
    <item seqNum="12" shotId="shot4781_35"/>
    <item seqNum="13" shotId="shot4781_39"/>
    <item seqNum="14" shotId="shot4781_40"/>
```

Figure 4. Concepts extraction

### 5.3. Concepts Correlation

The concepts used in the video data feature extraction process have a semantic relation between them. The measure of this relation is based on correlation formula between concepts. The aim of this correlation formula is to attribute a weight between a concept $C_i$ and the other concept. This weight is comprised between 0 and 1. 0 if there is not a relation between the two concepts and 1 if $C_1$ is a generalisation case of $C_2$. For example the weight between "Adult" and "Person" is 1 because an adult is a person. But the weight between "Adult" and "Male_Person" is 0,5.
Figure three illustrate the result of the application of the following correlate formula.

$$Cd = \frac{Number\ shot\ indexing\ A}{Number\ shot\ indexing\ A\ and\ B}$$

```xml
<?xml version="1.0" encoding="UTF-8"?>
<!DOCTYPE Indexing SYSTEM "index.dtd">
<Indexing>
    <Concept ConceptId="1" ConceptName="Actor">
        <SubConcept ConceptID="90" ConceptName="Person" Weight="1"/>
    </Concept>
    <Concept ConceptId="2" ConceptName="Adult">
        <SubConcept ConceptID="75" ConceptName="Male_Person" Weight="0,5012"/>
        <SubConcept ConceptID="90" ConceptName="Person" Weight="1"/>
        <SubConcept ConceptID="97" ConceptName="Reporters" Weight="0,5335"/>
        <SubConcept ConceptID="106" ConceptName="Single_Person" Weight="0,4901"/
    </Concept>
```

Figure5. Concepts correlation

### 5.4. Semantic Similarity distance

Semantic similarity or semantic relatedness is a concept whereby a set of documents or terms within term lists are assigned a metric based on the likeness of their meaning /semantic content. The determination of the degree of similarity between two semantic concepts is a difficult problem which arises in many applications. The similarity calculation between concepts can be based on the hierarchical link of specialization/generalization. The similarity between the documents thus returns to a simple mathematical calculation between the vectors which compose them.

We based the semantic similarity computation between two video shots on the correlation formula highlighted in the precedent section. Let us consider two shots A and B. "Person" and 'Airplane" are the concepts describing A. "Actor" and "Building" are the concepts describing B. The measure of the similarity between A and B is done by the summation of the weight according to the concepts composing the descriptor vector of the shot A ($V_A$) devising by the number concepts constituting $V_A$.

$$Sd(A,B) = \frac{\sum_{1}^{n} Cd(Va,Vb)}{n}$$

With Sd is the similarity distance, A and B are two different shots, Cd is the weight between concepts describing A by report B and n is the number of concept composing Vb.

The result of the semantic similarity distance is a semantic space, which is integrated in our work (Figure 6). The semantic distance is comprised between 0 and 1. The distance is value 1 if the two documents are similar and is value 0 if the two videos are completely different semantically. If a user looks at two video documents which have 1 as measure semantic distance can note easily that both videos indicate the same event. Both videos indicate the same event if they have the same concept vectors.

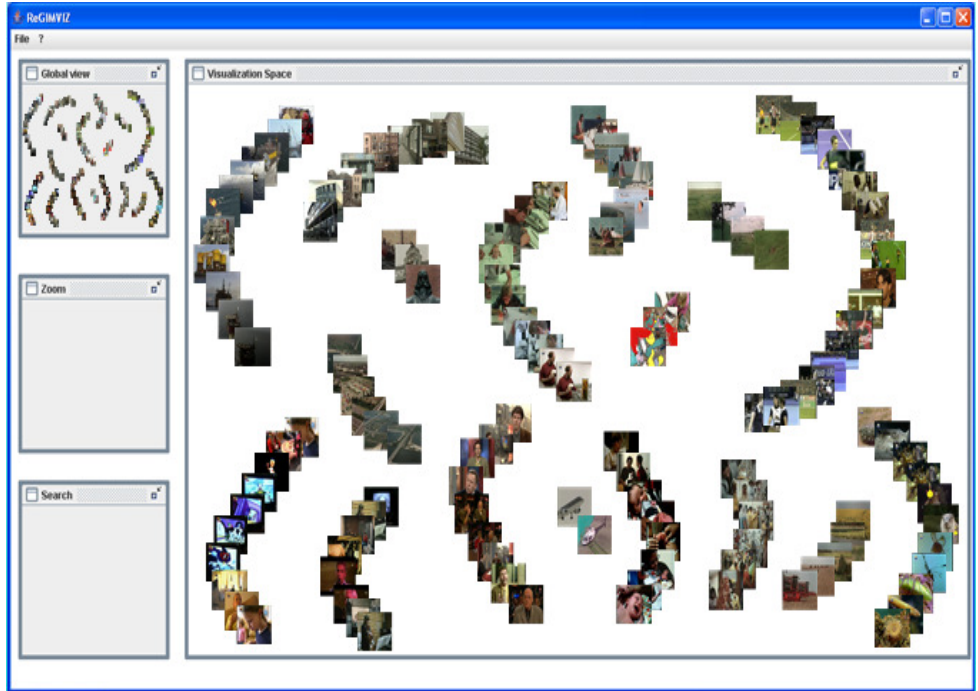

Figure6. Semantic Space

## 6. VISUALIZATION SYSTEM

The main goal of the visualization space is to provide an overview of available data to user. A 2D interface allowed visualizing the video data collection. Visualization is done by a projection of the classes obtained after the similarity computation.

In literature, several data visualization system based their representation on the use of metaphor. The use of the metaphor returns the interfaces more comprehensible and more meadows of thought of the user.

In our approach, the metaphor is inspired from the biological neuron operation for representing the collection in the visualization space. In the following section we explain the function of the biological neuron.

### 6.1. Neuron metaphor

A neuron, also known as a neurone or nerve cell, is an electrically excitable cell that processes and transmits information by electrical and chemical signalling. Chemical signalling occurs via synapses, specialized connections with other cells. Neurons connect to each other to form neural networks.

A typical neuron possesses a cell body (often called the soma), dendrites, and an axon. Dendrites are thin structures that arise from the cell body. An axon is a special cellular extension that arises from the cell body at a site called the axon hillock and travels for a distance. The cell body of a neuron frequently gives rise to multiple dendrites, but never to more than one axon.

Hence, the body represents the node (keyframe), the axon represents the relation between the classes and the dendrite represents the relation inter elements. In effects, the visualization graph is

International Journal of Computer Graphics & Animation (IJCGA) Vol.2, No.2/3, July 2012

in the form of neuronal network. The click on a keyframe is the stimulus which activates the others keyframe related to them.

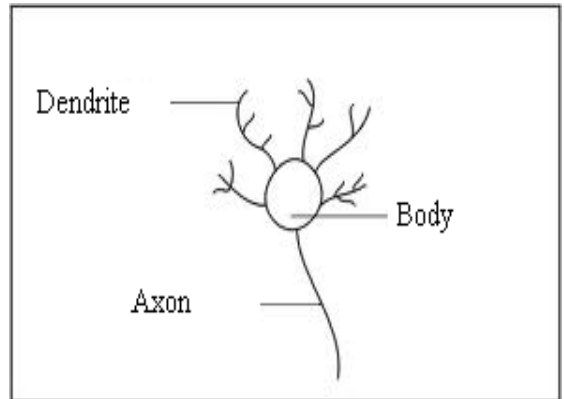

Figure7. Biological Neuron

## 6.2. Visualization graph

There is one conflicting requirements for visualizing large video data collection: The visualization space is limited. So, we need to show as many elements of interest as possible in such limited space. The network representation permits to decrease this problem.

The visualization graph is represented by keyframes and arcs. Arcs represent semantic relations between documents and between classes. The keyframe is the object representing the documents in the visualization space. Visualization in the form of network allows an effective browsing of video data collection. This is done by explicit the semantic relation between the classes and between the documents.

Visualization system posts a global overview of the corpus allowing an easy knowledge extraction of the corpora.  The problem is that the limited visualization space forbids the revelation of the totality of knowledge. So the documents should well be chosen to be posted. That always depends on the preferences and the expectations of users. The idea is to build user profile allowing visualization on demand.

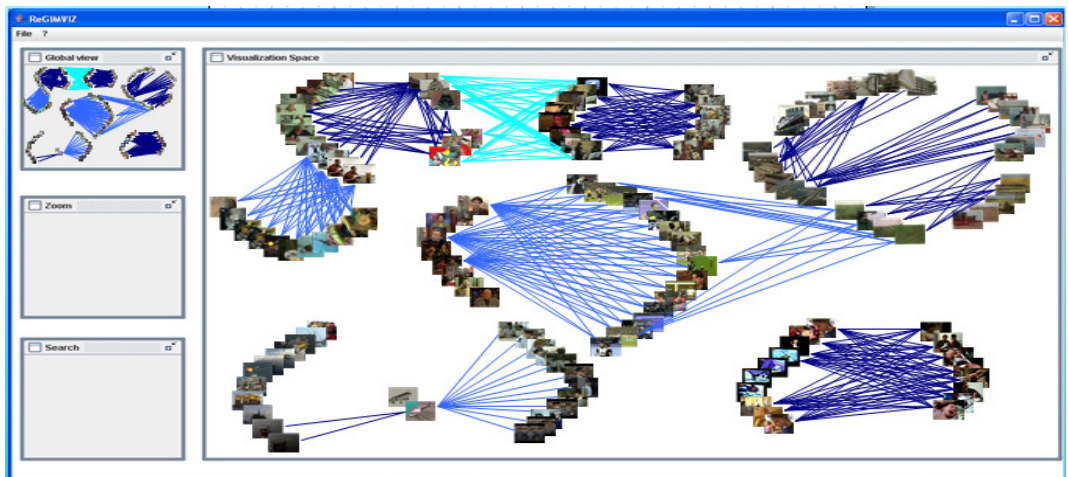

Figure8. Visualization Graph

## 7. PERSONALIZATION

Environment personalization consists in providing an appropriate environment to the user's profile. Profile is the regrouping of his/her preferences and his/her interest's centers. The personalization process is an incremental and interactive process in which the user is a decidor.
The creation of the user profile requires a collection of his preferences via its interaction with the system. Before his input, the system posts a global overview of the video collection. This overview reveals the possible maximum of topics existing in the video database.

### 7.1. User Interaction with the system

A user enters the system with an information need. The system posts an overview of the existing topic in the video collection. The user chooses the more important shot by clicking in the correspondent keyframe. This click is considered as stimulus which activates all the documents with an important relation with the choosing element. This activation is according by posting the interesting elements in the visualization space and eliminating the others. For example, when the user clicks on a keyframe corresponding to the topic of sport, the system will give us only the information's that are related to this topic (for example animals). In the topic of sport, if the user chooses football sequence, the documents which have an important relation with "football" must be posted on the visualization space. For the reason that we cannot predict the following action of the user, the system must keep a global overview of the corpus.

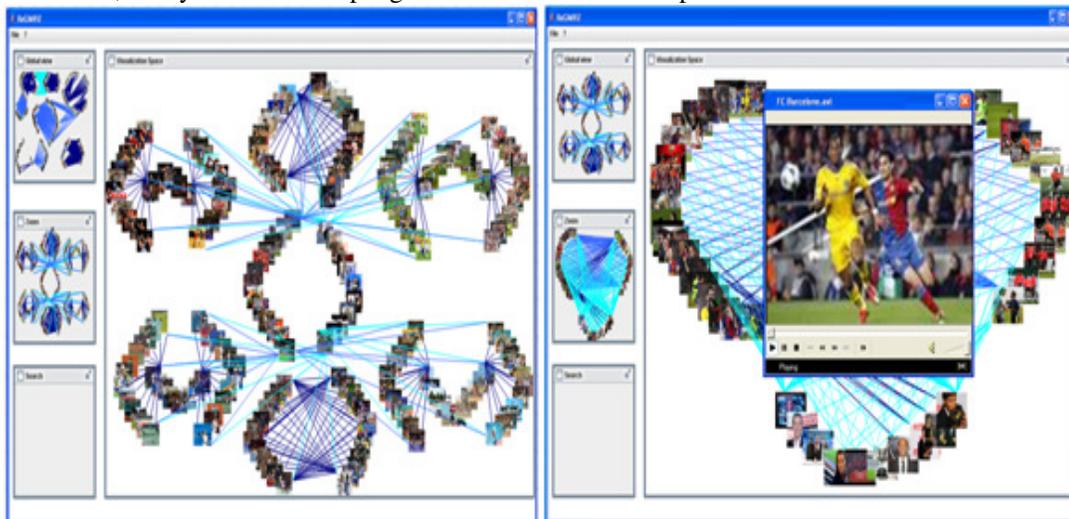

Figure9. Visualization space

### 7.2. User profile

The aim of user's profile is to simplify the exploration of the video data collection. For this, the knowledge interpretation must be easy to modify according to the user's preferences. So, we must weighting the edges connecting elements according to their importance for the user. Before, user input System doesn't have any idea about his preferences viewpoint. Hence, the construct of the user's profile needs the collection of his interest centers. This interest centers is done via the interaction between the user and the system.

At anytime of the exploration, the user can express his interest viewpoint by clicking the node of the network. Then, the system must modify the data representation by posting the most relevant shot from the video collection. This by assigning a weight user on the relation between the desired shot and the others shot composing the video collection.

$$R = Wu\ (Sd, Si)$$

Where R is the relation between the desired shot Sd and the shot Si, Wu is the user weight. Wu is comprised between 0 and 1.

The determination of the user's preference is extracted from various indicators. The static indicators describe the user personal information. The dynamic indicator which are extracted from the interaction with the system: navigation history, frequency consultation and duration consultation.

## 7. CONCLUSION AND FUTURE WORK

Visualization is an essential tool for exploring visual collections. To build a good visualization system, a set of related requirements should be taken into account. In this paper we have presented a video data visualization tool which covers the video data visualization process from the video feature extraction to the personalization of the information access. The principal objective of our tool is to simplify exploration, navigation and access to documents in large scale video corpora. That through a network visualization graph (Nodes and arcs coloured).

The performance of the currently integrated visualization approach still needs to be improved. The improvements will relate to the semantic classification part. Therefore, different semantic classification techniques have to be studied. We plan to pass to a 3D representation of the visualization graph. A long term goal of our work is to develop a generic video data retrieval system.

## REFERENCES

[1] Beecks .C, Skopal. T, Schoeffmann. K, & Seidl T,(2011) "Towards large-Scale Multimedia Exploration", 5th International Workshop on Ranking in Databases.
[2] Fan.J, Luo. H, & Elmagarmid. A. K,(2004) "Concept oriented indexing of video database toward more effective retrieval and browsing", IEEE Transactin on Image Processing, pp 974–992.
[3] AMARNATH. G, & RAMESH. J, (1997) "Visual information retrieval", Communications of the ACMMM .
[4] Snoek. C, Worring. M. Koelma. D, & Smeulders, (2006) "Learned lexicon-driven interactive video retrieval", In CIVR, pp 11.
[5] Campanella. M, Leonardi. R, & Migliorati, (2005), "Future viewer: an effcient framework for navigating and classifying audiovisual documents", In WIAMIS '05, Montreax, Switzerland.
[6] Campanella. M, Leonardi. R, & Migliorati, (2009), "Interactive visualization of video content and associated description for semantic annotation ", Journal In Signal, Image and Video Processing, Vol. 3, no. 2, pp. 183-196.
[7] Nguyen. G.P, & Worring. M, (2007), "Interactive access to large image collections using similarity based visualization", Journal of Visual Languages and Computing.
[8] Schoeffmann. K, Hopfgartner. F, Marques. O, Boeszoemenyi. L, & Jose. J.M, (2010), "Video browsing interfaces and applications: a review", SPIE reviews.
[9] Schaefer. G, (2011), "Image browsers Effective and efficient tools managing large image collections", IEEE International Conference on Multimedia Computing and Systems, pp 1-3.
[10] Hurst. W, Jarves. P, (2005), "Interactive, dynamic video browsing with the zoomslider interface", in IEEE ICME.

[11] Dragicevic. P, Ramos. G, Bibliowitcz. J, Nowrouzezahrai. D, Balakrishnan. R, & Singh. K, (2008), "Video browsing by direct manipulation", in SIGCHI on human factors in computing systems, pp 237-246.
[12] Shoeffmann. K, Boeszoermenyi, (2009), "Video browsing using interactive navigation summaries", In International Workshop on content-based Multimedia Indexing.
[13] Campanella. M, Leonardi. R, & Migliorati. P, (2005), "The future-viewer visual environment for semantic characterization of video sequences", In Proc. Of International Conference on Image Processing, pp 1209-1212.
[14] Worring. M, Snoek. C.G.M, de Rooij. O, Nguyen. G.P, & Smeulders W.M, (2007). "The Mediamill semantic video search engine". In IEEE ICASSP, Honolulu, Hawaii, USA.
[15] Snoek. C.G.M, et all (2009), "The mediamill TRECVID 2009 semantic video search engine". In Proc. TRECVID Workshop.
[16] Snoek. C.G.M, Freiburg. B, Oomen. J, & Ordelman. R, (2010), "Crowdsourcing rock n'roll multimedia retrieval". In Proc. ACM Multimedia.
[17] Snoek. C.G.M, (2010), "The mediamill Search Engine Video". In ACM Multimedia.
[18] Nguyen. G.P, & Worring. M, (2007), "Optimization of interactive visual similarity based search", In ACM TOMCCAP.
[19] Goeau. H, Thievre. J, Viaud,.M-L, & Pellerin. D, (2008), "Interactive visualization tool with graphic table of video contents". In CIVR.
[20] Del Fabro. M, Schoeffmann. K, & Boeszoermenyi. L, (2010), "Instant Video Browsing: A tool for fast Nonsequential Hierarchical Video Browsing". In Workshop of Intercative Multimedia Applications.
[21] Schoeffmann. K, Taschwer. M, & Boeszoemenyi. L, (2010), "The video explorer: a tool for navigation and searching within a single video based on fast content analysis". In ACM Multimedia Systems.
[22] Mûller. C, Smole. M, & Schoeffmann. K, (2012), "Demonstartion of a hierarchical Multi-layout 3D video browser". In IEEE Internatinal Conference on Multimedia and Expo.
[23] Schoeffmann. K, Ahlstrôm. D, & Bszormenyi. L, (2012), "Video browsing with a 3D thumbnail ring arranged by color similarity". In Advances in Multimedia Modeling, pp 639-641.
[24] Davenport. G, Smith. T.G.A, & Pincever. N, (1991), "Cinematic principles for multimedia", IEEE Computer Graphics & Applications, vol. 11, pp. 67–74,
[25] Brunelli. R, Mich. O, & Modena. C. M, (1999), "A survey on the automatic indexing of video data", Journal of Visual Communication and Image Representation, vol. 10, pp. 78–112.
[26] Snoek .C.G.M, Worring.M, (2008), "Concept-Based Video Retrieval", Foundation and trend in information retrieval, vol. 2, pp. 215–322.
[27] Karray .H and al (2008), "ReGIM at TRECVID 2008: high-level features extraction and video search", In TREC Video Retrieval Evaluation online proceedings.
[28] Datta. R, Joshi. D, Li. L, & Wang .J.Z, (2008), "Image retrieval: Ideas, influences and trends of the new age", ACM Computing Surveys, vol. 40, pp. 1–60.
[29] M. S. Lew, ed. (2001), "Principles of Visual Information Retrieval". Springer.
[30] Smeulders. A.W.M, Worring. M, Santini. S, Gupta. A, & Jain. R,(2000), "Contentbased image retrieval at the end of the early years", IEEE Transactions on Pattern Analysis and Machine Intelligence, vol. 22, pp. 1349–1380.
[31] Tuytelaars. T, & Mikolajczyk. K,(2008), "Local invariant feature detectors: A survey", Foundations and Trends in Computer Graphics and Vision, vol. 3, pp. 177–280.
[32] Van deSande. K, Gevers. T, & Snoek. C.G.M, (2010), "Evaluating Color Descriptors for Object and Scene Recognition", IEEE Transactions on Pattern Analysis and Machine Intelligence, vol. 32.
[33] Carson. C, Belongie. S, Greenspan. H, & Malik. J,(2002), "Blobworld: Image segmentation using expectation-maximization and its application to image querying", IEEE Transactions on Pattern Analysis and Machine Intelligence, vol. 24, pp. 1026–1038.
[34] Deng. Y, & Manjunath .B.S,(2001), "Unsupervised segmentation of color-texture regions in images and video", IEEE Transactions on Pattern Analysis and Machine Intelligence, vol. 23, pp. 800–810.

International Journal of Computer Graphics & Animation (IJCGA) Vol.2, No.2/3, July 2012

[35] Hoang. M.A, Geusebroek. J.M, & Smeulders. A.W.M, (2005), "Color texture measurement and segmentation", Signal Processing, vol. 85, pp. 265–275.
[36] Gevers. T, (2002), "Adaptive image segmentation by combining photometric invariant region and edge information", IEEE Transactions on Pattern Analysis and Machine Intelligence, vol. 24, pp. 848–852.
[37] Van Germert .J.C, Geusedroek .J.M, Snoek .C.G.M, Worring .M & Smeulders .A.W.M, (2006), "the challenge problem for automated detection of 101 semantic concepts in multimedia", In proceeding of the ACM International Conference on Multimedia, pp. 421–430.
[38] Byrne. D, Doherty. A.R, Snoek .C.G.M, (2010), "Everyday concept detection in visual lifelogs: validation, relationships and trends", In proceeding of Multimdia Tools Applications, pp 119–144.

**Authors**

**Jamel SLIMI** is a phd student in the National Engineering School of Sfax, Tunisia. Member of the IEEE student Brench in Sfax, Tunisia. Member of the IEEE computer society. Member of the ReGIM Laboratory (Research Groups on Intelligent Machines). 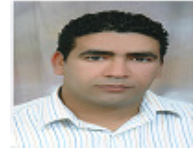

**Anis Ben AMMAR** is assistant professor in Higher Institute of Business Administration of Sfax, Tunisia. Member of the IEEE computer society. Member of the ReGIM Laboratory (Research Groups on Intelligent Machines). 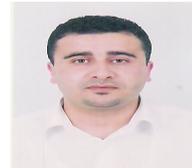

**Adel M.ALIMI** is professor in National Engineering School of Sfax, Tunisia. Member of the IEEE computer society. President of the ReGIM Laboratory (Research Groups on Intelligent Machines). 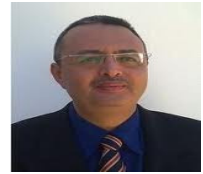